\documentclass[aps, prd, twocolumn, notitlepage, nofootinbib]{revtex4-1}
\usepackage{graphicx}
\usepackage{float}
\usepackage{amsmath}
\usepackage{color}
\usepackage{tensor}
\usepackage{epstopdf}
\usepackage{xcolor}
\usepackage{wasysym}
\usepackage{hyperref}
\hypersetup{
    colorlinks,
    linkcolor={blue!50!black},
    citecolor={red!50!black},
    urlcolor={blue!80!black}
}

\usepackage{multibib}
\newcommand{\be}{\begin{equation}}
\newcommand{\ee}{\end{equation}}
\newcommand{\ba}{\begin{eqnarray}}
\newcommand{\ea}{\end{eqnarray}}

\newcommand{\beq}{\begin{equation}}
\newcommand{\eeq}{\end{equation}}
\newcommand{\beqa}{\begin{eqnarray}}
\newcommand{\eeqa}{\end{eqnarray}}
\newcommand{\nn}{\nonumber}

\begin{document}

\title{Gravitational Lensing by Black Holes in Einsteinian Cubic Gravity}

\author{Mohammad Bagher Jahani Poshteh}
\email{mb.jahani@iasbs.ac.ir }
\affiliation{Department of Physics and Astronomy, University of Waterloo, Waterloo, Ontario, Canada, N2L 3G1}
\affiliation{Department of Physics, Institute for Advanced Studies in Basic Sciences (IASBS), Zanjan 45137-66731, Iran}

\author{Robert B. Mann}
\email{rbmann@uwaterloo.ca}
\affiliation{Department of Physics and Astronomy, University of Waterloo, Waterloo, Ontario, Canada, N2L 3G1}
\affiliation{Perimeter Institute, 31 Caroline St. N., Waterloo,
	Ontario, N2L 2Y5, Canada}

%
%\date{\today}


\begin{abstract}
We investigate the predictions of Einsteinian Cubic Gravity (ECG) for the lensing effects due to supermassive black holes at the center of Milky Way and other galaxies. Working in the context of spherical symmetry, we obtain the metric function from a continued fraction method and find that both time delays and the angular positions of images considerably deviate from general relativity, as large as milliarcseconds. This suggests that observational tests of ECG are indeed feasible.
\end{abstract}

\maketitle

\section{Introduction}

The deflection of light near a massive body is one of the main implications of general relativity (GR) and its investigation  in the weak field regime has a history of nearly one century~\cite{Dyson,Einstein,Schneider}. In the strong field regime the study of gravitational lensing traces back to the work of Darwin~\cite{Darwin59,Darwin61}, who studied the deflection of light near the Schwarzschild black hole. Lensing effects in strong gravitational fields were resurrected in~\cite{Ellis}, where the authors investigated the infinite number of images on each side of the optical axis of a Schwarzschild black hole and called them {\it relativistic images}. It was later shown~\cite{Keeton} that it is not necessary for the lens to be a black hole to produce relativistic images; rather, any sufficiently compact object could create relativistic images. A detailed study of relativistic images presented in~\cite{Virbhadra} showed that the time delay between the outermost two relativistic images could be used to obtain a very accurate value of the mass of the black hole. Furthermore, given the mass, the angular separation between relativistic images would give us the distance to the black hole.

In this paper we investigate the gravitational lensing (GL) of black holes in Einsteinian cubic gravity (ECG).
ECG is the unique cubic theory of gravity that shares its graviton spectrum with Einstein gravity and has a dimension-independent coupling constant~\cite{Pablo1}.  The Lagrangian density of this theory is given by
\begin{equation}
\label{lagden}
{\cal L} = \frac{1}{2 \kappa} \left[-2 \Lambda + R \right] + \beta_1 \chi_{4} + \kappa \left[\beta_2 \chi_{6} + \lambda {\cal P} \right],
\end{equation}
where $\chi_{4}$ and $\chi_{6}$ are, respectively, four- and six-dimensional Euler densities and correspond to the usual Lovelock terms, and
\begin{align}
{\cal P} = 12 \tensor{R}{_\mu ^\rho _\nu ^\sigma} \tensor{R}{_\rho ^\gamma _\sigma ^\delta}\tensor{R}{_\gamma ^\mu _\delta ^\nu} + R_{\mu\nu}^{\rho\sigma} R_{\rho\sigma}^{\gamma\delta}R_{\gamma\delta}^{\mu\nu}
\nn\\
 - 12 R_{\mu\nu\rho\lambda}R^{\mu \rho}R^{\nu\sigma} + 8 R_\mu^\nu R_\nu ^\rho R_\rho ^\mu .
\end{align}
Up to cubic order in curvature, ECG is the most general theory of gravity that admits a static spherically symmetric solution characterized by a single metric function~\cite{Pablo2}. In (3+1) dimensions the terms proportional to  $\beta_1$ and $\beta_2$ have no effect on the field equations.   However the  new term ${\cal P}$ does, and retains the interesting properties noted above. ECG is thus singled out as a unique generalization of Einstein gravity in 3 spatial dimensions with terms cubic in the curvature but possessing only a single coupling constant.

For these reasons we regard ECG as a phenomenological competitor to GR that merits further study. Recently, we have obtained an approximate analytical solution to ECG in four dimensions by employing continued fraction method \cite{OURS}. This solution holds everywhere outside the horizon and can be used in the same way as analytical solution. In this paper we use the continued fraction solution of \cite{OURS} to study the gravitational lensing in ECG, and investigate its observational signature.

We note that GL effects have been studied for many different black holes in GR and alternative theories~\cite{Torres,Bozza2003,Bhadra,Whisker,Eiroa,Jing2009,Wei,Cai,Jing2017}, with the strong field limit approximation of Bozza~\cite{Bozza2002,Bozza2001} employed throughout. Although useful, Bozza's analytical treatments have been criticized for their accuracy~\cite{Virbhadra}. We shall therefore use the numerical method of~\cite{Ellis} to study GL by black holes in  ECG.

Our most interesting finding is that the difference between the angular positions of primary and secondary images in ECG and GR could be as large as milliarcseconds.
Also, the predicted values of time delay between these images are different in GR and ECG, and the difference could be as large as seconds.
These suggest that observational tests of ECG are indeed feasible.

The outline of our paper is as follows.
 In the next section we   review our continued fraction method to find the approximate analytic spherically symmetric solution to ECG.
  In Sec.~\ref{sec:bhlensing} we use the Lagrangian of massless particle to obtain equations needed to investigate the lensing effects. This would include the relation for the bending angle, time delay and magnification of images. We use these equations in Sec.~\ref{sec:sgrlensing} to study GL of SMBHs, for  Sgr A* and those at the centers of 13 other galaxies. We conclude our paper in Sec.~\ref{sec:con}.
We will work in units where $G = c= 1$.

\section{Black Hole Solution in Einsteinian Cubic Gravity}
\label{sec:bhinecg}

In this section we briefly review the continued fraction solution for the metric function of ECG obtained in \cite{OURS}. We restrict ourselves to asymptotically flat, static and spherically symmetric vacuum black holes with the line element
\begin{equation}\label{eqn:le}
ds^2= -f(r)dt^2+\frac{dr^2}{f(r)}+r^2(d\theta^2+\sin^2\theta d\phi^2).
\end{equation}
Substitute this metric into the Lagrangian~\eqref{lagden}, the field equation for Einsteinian cubic gravity reads~\cite{Mann1}
\begin{align}\label{eqn:feq}
&-(f-1)r - \lambda \bigg[\frac{f'^3}{3} + \frac{1}{r} f'^2 - \frac{2}{r^2} f(f-1) f' 
\nn\\
&- \frac{1}{r} f f'' (rf' - 2(f-1)) \bigg] = 2 M,
\end{align} 
where $f$ stands for $f(r)$ and a prime denotes differentiation with respect to $r$. The constant of integration  $M$ appearing on the right-hand side of \eqref{eqn:feq} can be shown to be the physical mass of the black 
hole \cite{Pablo1,Hennigar:2017ego,Mann1}. Also, we will assume $\lambda > 0$ in what follows.  

Consider the near horizon series expansion of the metric function: 
\be\label{eqn:nh_expand} 
f_{\rm nh}(r) = 4 \pi T (r-r_+) + \sum_{n=2}^{n = \infty} a_n (r-r_+)^n \, ,
\ee 
which ensures that the metric function vanishes linearly at the horizon ($r=r_+$), and $T = f'(r_+)/4\pi$ is the Hawking temperature.  By substituting this ansatz into the field equations~\eqref{eqn:feq}, we can find the temperature and mass in terms of $r_+$ and the coupling $\lambda$:
\begin{align}\label{eqn:mass_temp}
M &= \frac{r_+^3}{12 \lambda^2} \left[r_+^6 + (2 \lambda - r_+^4) \sqrt{r_+^4 + 4 \lambda} \right] \, ,
\nn\\
T &= \frac{r_+}{8 \pi \lambda} \left[ \sqrt{r_+^4 + 4 \lambda} - r_+^2 \right] \, .
\end{align}   
One then finds that $a_2$ is left undetermined by the field equations, while all $a_n$ for $n > 2$ are determined by (rather messy) expressions involving $T$, $M$, $r_+$, and $a_2$.  

The asymptotic solution to \eqref{eqn:feq}  is~\cite{OURS,Hennigar:2017ego}  
\begin{equation}
f(r) \approx 1 - \frac{2 M}{r}-\frac{36 \lambda M^2 }{r^6} + \frac{184}{3} \frac{\lambda M^3}{r^7} + \mathcal{O}\left(\frac{M^3 \lambda^2}{r^{11}} \right),
\end{equation}
and
to bridge the gap between this solution and the near horizon approximation, one can numerically solve the equations of motion in the intermediate regime. This is done by picking, for a given value of $M$ and $\lambda$,  a value for $a_2$ and using it in the near horizon expansion to obtain the initial data
\begin{align}\label{eqn:nh_data}
f(r_+ + \epsilon) &= 4 \pi T \epsilon + a_2 \epsilon^2 \, ,
\nn\\
f'(r_+ + \epsilon) &= 4 \pi T  + 2 a_2 \epsilon \, ,
\end{align}
where $\epsilon$ is some small, positive quantity.   A satisfactory solution is the one that agrees with the asymptotic expansion at a sufficiently large distance from the black hole. 

In practice we find that this only happens for a unique value of $a_2$ which we denoted by $a_2^\star$ \cite{OURS}. By fitting the numerical results we   find
\begin{equation}
\label{eqn:approx_a2} 
a_2^\star \left(z= \lambda / M^4 \right) \approx -\frac{1}{M^2} \frac{1 + 2.1347 z + 0.0109172 z^2}{4 + 15.5284 z + 8.03479 z^2},
\end{equation}
which is accurate to three decimal places or better in the interval $\lambda/M^4 \in [0, 5]$.

Now, to obtain an analytic solution with the continued fraction method, we first compactify the spacetime interval outside of the horizon by using the coordinate $x=1-r_+/r$, and rewriting the metric function as
\begin{equation}
\label{eqn:cfrac_ansatz} 
f(x) = x \left[1 - \varepsilon(1-x) + (b_0 - \varepsilon)(1-x)^2 + \tilde{B}(x)(1-x)^3 \right],
\end{equation}
where
\begin{equation}
\tilde{B}(x) = \cfrac{b_1}{1+\cfrac{b_2 x}{1+\cfrac{b_3 x}{1+\cdots}}} 
\end{equation}
is a continued fraction whose coefficients are to be determined from the field equations. Substituting the asymptotic (near $x=1$) expansion of~\eqref{eqn:cfrac_ansatz} into the field equation~\eqref{eqn:feq} yields
\begin{align}
\varepsilon &= \frac{2 M}{r_+} - 1,  \qquad b_0 = 0 \, .
\end{align}
Next, expanding~\eqref{eqn:cfrac_ansatz} near the horizon ($x=0$), the remaining coefficients can be fixed in terms of $T$, $M$, $r_+$ and one free parameter, $b_2$.  We find
\be 
b_1 = 4 \pi r_+ T + \frac{4 M}{r_+} - 3,
\ee
while $b_2$ is related to the coefficient $a_2$ appearing in the near horizon expansion~\eqref{eqn:nh_expand} by
\be\label{eqn:frac_b2}
b_2 = - \frac{r_+^3 a_2 + 16 \pi r_+^2 T + 6(M-r_+)}{4 \pi r_+^2 T + 4 M - 3 r_+} \, .
\ee
All higher order coefficients are then determined by the field equations in terms of $T$, $M$, $r_+$ and $b_2$ (or, equivalently, $a_2$). Since $b_2$ is not fixed by the field equations its value must be manually input into the continued fraction.  The appropriate thing to do is to use the value of $a_2^\star$ (as determined through the numerical method) in Eq.~\eqref{eqn:frac_b2}.  While the numerical integration of the field equations is very sensitive to the precision with which $a_2^\star$ is specified, the continued fraction is much less so, and a good approximation is obtained even with just a few accurate digits.

\section{Black hole lensing}
\label{sec:bhlensing}

\begin{figure}
	\centering
	\includegraphics[width=0.5\textwidth]{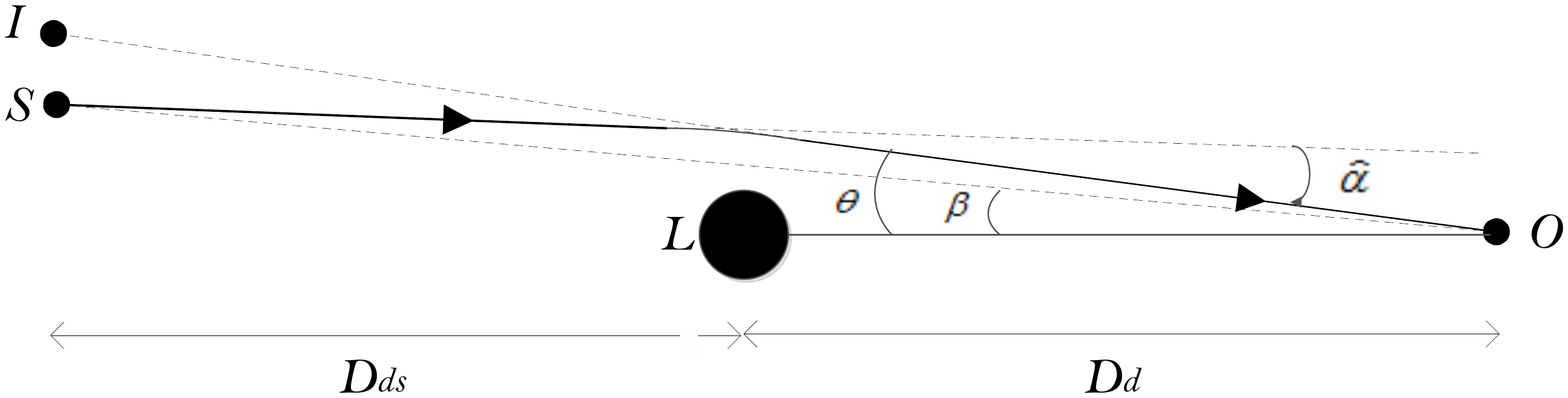}
	\caption{{\bf The lens diagram}: As the light ray pass the black hole it deflects by an angle $\hat{\alpha}$. Those rays which pass closer to the black hole would have a larger deflection angle. If $\hat{\alpha}>2\pi$, the corresponding light ray winds the black hole at least once, before reaching the observer. These rays would make the relativistic images. Here $S$, $I$, $O$, and $L$ stand, respectively, for the source, image, observer, and the lens which is a black hole in our study. $\beta$ is the actual angular position fo the source w.r.t. the line of sight to the black hole. $\theta$ is the angular position of the image. $D_d$ and $D_{ds}$ represent, respectively, the distance from lens to observer and from lens to the source.}
	\label{fig:lens_diag}
\end{figure}

In this section we obtain some basic equations needed to study gravitational lensing by black holes. For the line element (\ref{eqn:le}), the Lagrangian is given by
\be\label{eqn:lag}
2\mathcal{L}=g_{\mu\nu}\dot{x}^\mu\dot{x}^\nu=-f\dot{t}^2+\frac{\dot{r}^2}{f}+r^2\sin^2(\vartheta)\dot{\phi}^2.
\ee
We assume that the observer, black hole, and the source lie on the equatorial plane $\vartheta=\pi/2$. We can then write the constants of motion as
\be
E=-\frac{\partial \mathcal{L}}{\partial \dot{t}}=f\dot{t},   \qquad L_z=-\frac{\partial \mathcal{L}}{\partial \dot{\phi}}=-r^2\dot{\phi}.
\ee
For null geodesics we have $\mathcal{L}=0$, and Eq. (\ref{eqn:lag}) can be written in the following form
\be\label{eqn:lagf1}
\frac{1}{fr^2}\left(\frac{dr}{d\phi}\right)^2=\dfrac{r^2}{f}\frac{E^2}{L_z^2}-1.
\ee
At the radius of   closest approach $r=r_0$, we have $\frac{dr}{d\phi}=0$, so from the r.h.s. of Eq. (\ref{eqn:lagf1}) we find $E^2/L_z^2=f_0/r_0^2$, where $f_0$ is the value of the metric function at $r=r_0$. Then, we can write Eq. (\ref{eqn:lagf1}) as
\be\label{eqn:dphidr}
\frac{d\phi}{dr}=\frac{1}{r\sqrt{\left(\frac{r}{r_0}\right)^2f_0-f}}.
\ee
A schematic diagram of the lensing effect is presented in Fig.~\ref{fig:lens_diag}. $D_d$ and $D_{ds}$ represent, respectively, the distance of the lens (L) from the observer (O) and the source (S). We assume that $D_d, D_{ds}\gg r_0$, so, we can write the deflection angle as~\cite{weinberg1972}
\be\label{eqn:alpha_hat}
\hat{\alpha}(r_0)=2\int_{r_0}^{\infty}\frac{dr}{r\sqrt{\left(\frac{r}{r_0}\right)^2f_0-f}}-\pi.
\ee

Now, let us write Eq. (\ref{eqn:lag}) for the null geodesic in the form:
\be
\frac{1}{f^2}\left(\frac{dr}{dt}\right)^2=1-\dfrac{f}{r^2}\frac{L_z^2}{E^2}.
\ee
Since $\frac{dr}{dt}=0$ at $r=r_0$ we obtain
\be\label{eqn:dtdr}
\frac{dt}{dr}=\frac{1}{f\sqrt{1-\left(\frac{r_0}{r}\right)^2\frac{f}{f_0}}}.
\ee
The difference between the time for the photons to travel the physical path from the source to the observer and the time it takes to reach the observer in flat spacetime, i.e. when there is no black hole between the source and the observer, is referred to as the time delay. Using Eq. (\ref{eqn:dtdr}) the time delay of an image is given by
\be\label{eqn:time_delay}
\tau(r_0)=\left[\int_{r_0}^{r_s}dr+\int_{r_0}^{r_o}dr\right]\frac{1}{f\sqrt{1-\left(\frac{r_0}{r}\right)^2\frac{f}{f_0}}}-D_s\sec\beta,
\ee
where  $D_s=D_d+D_{ds}$ is the distance from observer to the source, $r_s=\sqrt{D_{ds}^2+D_s^2\tan^2\beta}$, and $r_o=D_d$, with $\beta$   the angular position of the source.

The image angular position, $\theta$, obeys the following relation
\be\label{eqn:lens_eq}
\tan\beta=\tan\theta-\mathcal{D}\left[\tan\theta+\tan(\hat{\alpha}-\theta)\right],
\ee
which is known as the lens equation~\cite{Ellis},  where $\mathcal{D}=D_{ds}/D_s$. The impact parameter is given by~\cite{Virbhadra98}
\be\label{eqn:impact}
J=\frac{r_0}{\sqrt{f_0}}=D_d\sin\theta,
\ee
and the image magnification by
\be\label{eqn:maggi}
\mu=\left(\frac{\sin\beta}{\sin\theta}\frac{d\beta}{d\theta}\right)^{-1}.
\ee
To find the magnification, $\mu$, we need the first derivative of the deflection angle w.r.t. $\theta$
\be\label{eqn:alpha_theta}
\frac{d\hat{\alpha}}{d\theta}=\frac{d\hat{\alpha}}{dr_0}\frac{dr_0}{d\theta}.
\ee
Given the metric function, we can obtain $\frac{dr_0}{d\theta}$ from Eq. (\ref{eqn:impact}). The derivative $\frac{d\hat{\alpha}}{dr_0}$ is a bit tricky. Let us write the deflection angle as
\be
\hat{\alpha}(r_0)=2\int_{r_0}^{C}\frac{dr}{r\sqrt{\mathcal{F}}}-\pi,
\ee
where $\mathcal{F}=\left(\frac{\textsf{t}r}{r_0}\right)^2f_0-f(\textsf{t} r)$. We will take $\textsf{t} \rightarrow 1$ and $C\rightarrow\infty$ at the end of our calculations. Using the Leibniz integral rule we find
\be\label{eqn:alpha_leibniz}
\frac{d\hat{\alpha}(r_0)}{dr_0}=-2\frac{1}{r_0\sqrt{\textsf{t} f_0-f(\textsf{t} r_0)}}+2\int_{r_0}^{C}\frac{\partial}{\partial r_0}\left(\frac{1}{r\sqrt{\mathcal{F}}}\right)dr,
\ee
in which the second term can be written as
\be
- \int_{r_0}^{C}\frac{1}{r \mathcal{F}^{3/2}}\frac{\partial \mathcal{F}}{\partial r_0}dr=2\int_{r_0}^{C}\frac{1}{r}\frac{\partial}{\partial r}\left(\frac{1}{\sqrt{\mathcal{F}}}\right)\frac{\partial\mathcal{F}}{\partial r_0}\frac{\partial r}{\partial\mathcal{F}}dr, \nn
\ee
which, by integrating by parts, gives a term that cancels the first term in Eq. (\ref{eqn:alpha_leibniz}) at the limits $\textsf{t} \rightarrow 1$ and $C\rightarrow\infty$. We are then left with
\be\label{eqn:alpha_r}
\frac{d\hat{\alpha}(r_0)}{dr_0}=-2\int_{r_0}^{\infty}\frac{dr}{\sqrt{\mathcal{F}}}\frac{\partial \tilde{\mathcal{F}}}{\partial r},
\ee
with
\be
\tilde{\mathcal{F}}=\frac{1}{r}\frac{\partial\mathcal{F}}{\partial r_0}\frac{\partial r}{\partial\mathcal{F}}.
\ee
In the following sections we use these results to study the gravitation lensing effects by black holes in general relativity as well as ECG.

\section{Lensing by supermassive black holes}
\label{sec:sgrlensing}

In this section we study the lensing effects by the supermassive black holes (SMBHs) at the center of the Milky Way  and 13 other galaxies.
Our aim is to compare the lensing predictions of GR with those of ECG. Using the metric functions for both GR and ECG, we numerically solve equations (\ref{eqn:alpha_hat}), (\ref{eqn:lens_eq}), (\ref{eqn:maggi}), and (\ref{eqn:time_delay}),  to respectively find their deflection angles, angular positions of their images, their magnifications, and their time delays. Lensing by Sgr A* in GR has been extensively studied in \cite{Ellis,Keeton,Virbhadra} by numerical methods. Here we recalculate GR results for the updated values of the mass of Sgr A* $M=5.94\times 10^9 \, {\rm m}$ and the distance $D=2.43\times 10^{20} \, {\rm m}$ from Earth~\cite{mnd}.

To find the results of ECG, we have used the metric function \eqref{eqn:cfrac_ansatz} obtained by the continued fraction method \cite{OURS}. We have furthermore constrained the coupling constant of ECG not to be larger than $\lambda = 4.57 \times 10^{22} M_{\astrosun}^4$; with this value, ECG passes all the Solar System tests to date \cite{OURS}. Assuming  the largest possible value of $\lambda$ allowed by Solar System tests, we find that the lensing effects from ECG  differ significantly from the GR predictions.

In Table \ref{tab:Tablei}, by using Eqs. (\ref{eqn:alpha_hat}) and (\ref{eqn:lens_eq}), we have calculated the bending angle $\hat{\alpha}$ and the angular image position $\theta$ for images on the same side as the source and on the opposite side of it, which are known as primary and secondary images, respectively. We have taken ${\cal D}=0.5$; meaning that the lens-source distance is the same as the lens-observer distance. The results are presented both in the case of GR and ECG with the coupling constant $\lambda/M_{Sgr A*}^{4}\approx 1.76\times 10^{-4}$. One could see that the ECG results for deflection angle and image angular positions ($\theta_p$ or $|\theta_s|$) are less than their corresponding values in GR.  
 
We have previously shown \cite{OURS} that ECG, with  coupling constant $\lambda/M_{Sgr A*}^{4}\approx 1.76\times 10^{-4}$, would enlarge the shadow of Sgr A* by an amount less than 1 nanoarcsecond. This is far lower than the resolution of today's observational facilities such as Event Horizon Telescope \cite{eht,Akiyama}
and occurs because the size of the shadow of Sgr A* is of order of $10^{-5}$ arcseconds  whether or not its gravitational field is governed by GR or ECG. The difference between GR and ECG results for the shadow size is three orders of magnitude smaller  and is about 1 nanoarcsecond.

However we have shown here that the situation is not quite so grim: the 
difference between the angular positions of primary/secondary images in ECG (with the same value of $\lambda$) and GR could be of order of miliarcseconds.
This is due to the fact that, for the source positions that we considered here, although the angular positions of primary/secondary images in GR or ECG are of the order of arcseconds,  the difference between the GR and ECG results can be as large as a few milliarcseconds, and so are feasibly distinguishable with present or near-future observations.

In Table \ref{tab:Tableii}, we have obtained the magnification $\mu$ of the primary and secondary images of Table \ref{tab:Tablei} by using Eqs. (\ref{eqn:maggi}), (\ref{eqn:alpha_theta}), and (\ref{eqn:alpha_r}); the time delay $\tau$ of the primary images have been calculated by using Eq. (\ref{eqn:time_delay}). We have not shown  explicit results for the secondary images, but have instead given the difference $t_d=\tau_s-\tau_p$ between the time delay of the secondary and the primary images, the so called differential time delay, since it is of more observational importance. 
 
Suppose that the source is pulsating. Every phase in its period would then appear in the secondary image, $t_d$ minutes after it appears in the primary image. Comparing the results of GR and ECG in Table \ref{tab:Tableii}, it is obvious that the differential time delay $t_d$ is lower if ECG correctly describes the strong gravitational field near the black hole. ECG, in addition, decrease the magnifications $\mu_p$ and $|\mu_s|$ by a small amount.

Of course observationally it is the images that are detected and not the source itself.  While it is possible under certain circumstances to to find the distance $D_{ds}$ to the source from its redshift~\cite{Schneider}, the angular position $\beta$ is not directly observable. In what follows, we propose a scheme to find $\beta$ from the primary and secondary image positions and their differential time delays, assuming  $D_{ds}$ is known, along with the mass of the lens.
\begin{figure}[htp]
	\centering
	\includegraphics[width=0.45\textwidth]{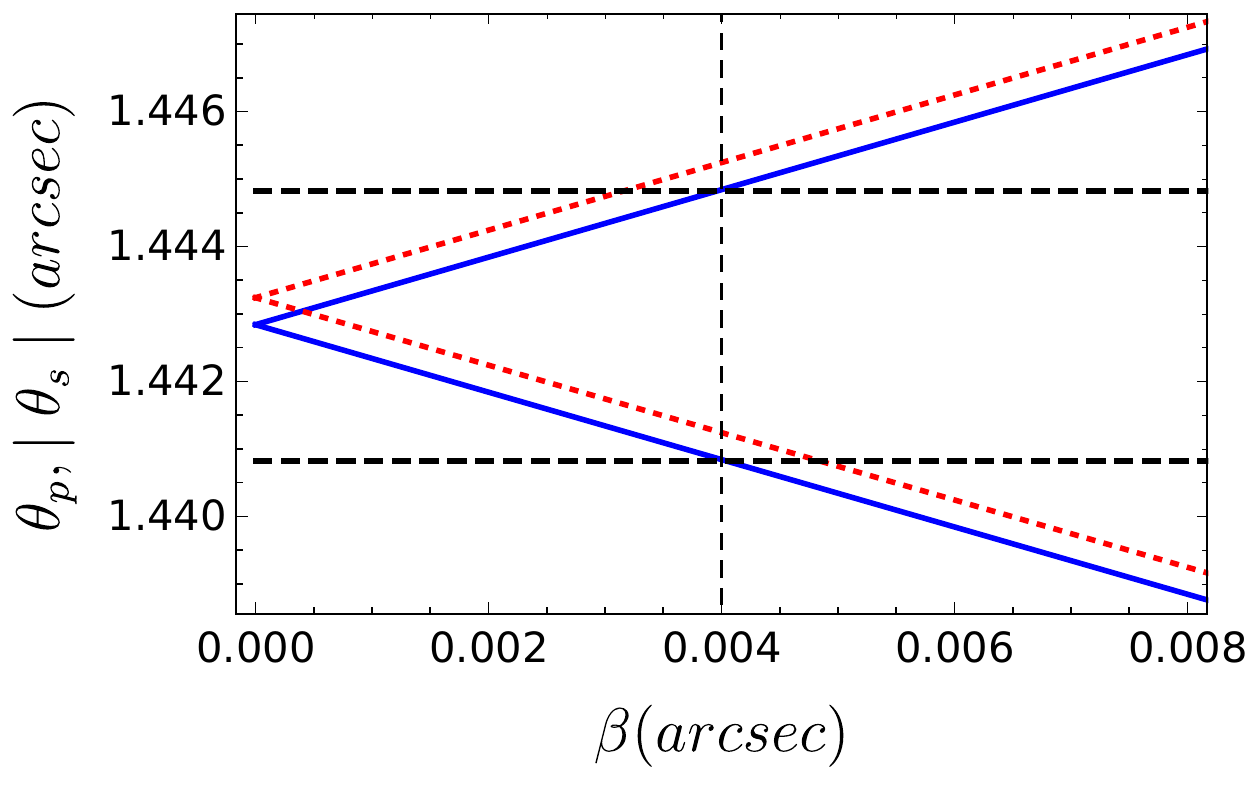}
	\includegraphics[width=0.45\textwidth]{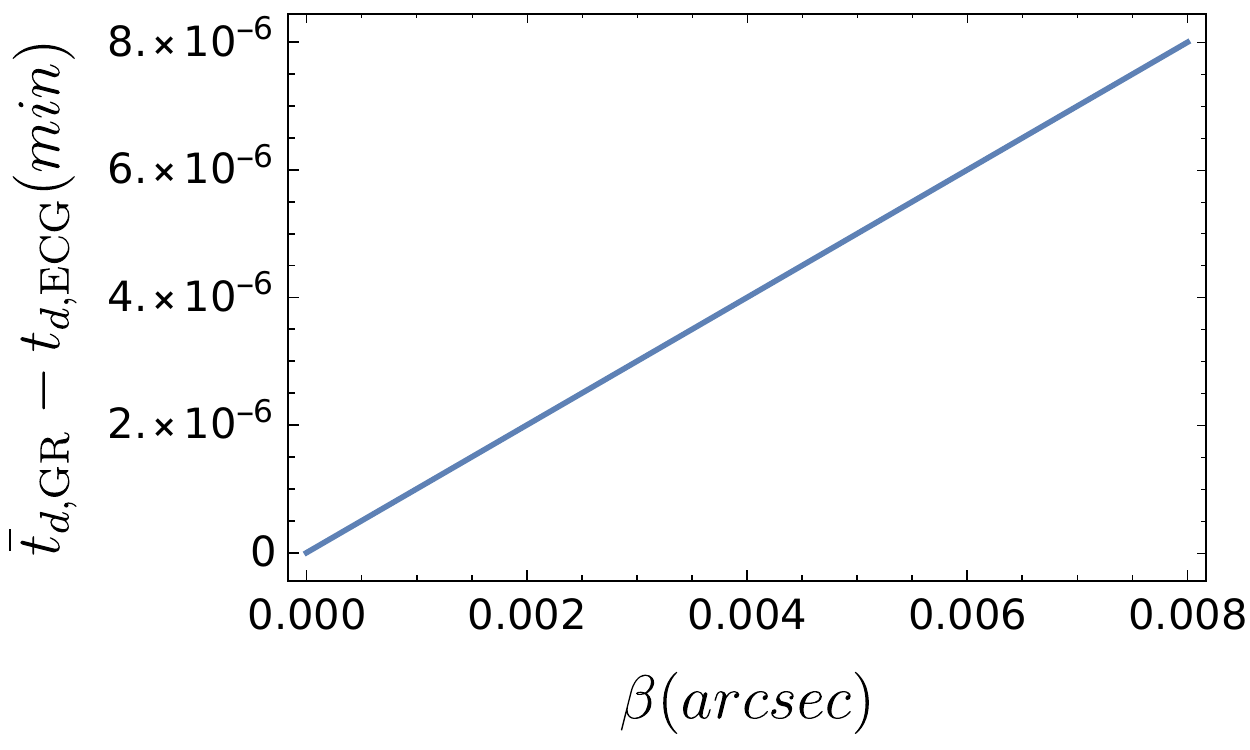}
	\caption{\textbf{Finding the source position}: \textit{Top}: Image positions as a function of the angular source position $\beta$ in GR (dotted, red curve) and ECG (solid, blue curve) with $\mathcal{D}=0.5$.  Those lines with positive slope correspond to the primary image position $\theta_{p}$ and those with negative slope to the secondary image position $|\theta_{s}|$. The horizontal dashed black lines indicate the position of the primary (upper line) and secondary (lower line) images in a particular observation. Each of these horizontal lines crosses both the dotted red and solid blue lines; $\beta$ is determined by finding a common intersection point for the two curves, illustrated by the vertical dashed black line. We have used Sgr A* as the lens with $M_{Sgr A*}=5.94\times 10^9 \, {\rm m}$ and $D_d=2.43\times 10^{20} \, {\rm m}$, and have taken $\lambda/M_{Sgr A*}^{4}\approx 1.76\times 10^{-4}$.
	\textit{Bottom}: Difference between the differential time delay in GR with $\mathcal{D}=0.49972$, $\bar{t}_{d,{\rm GR}}$, and that in ECG with $\mathcal{D}=0.5$, $t_{d,{\rm ECG}}$.
	}
	\label{fig:finding_source}
\end{figure}

In the top plot of Fig.~\ref{fig:finding_source} we have plotted   $\theta_{p}$ and $|\theta_{s}|$, the respective primary and secondary angular image positions (depicted by dashed horizontal lines) in GR and ECG with $\mathcal{D}=0.5$.  Each of these lines crosses both the plot of GR and ECG. We do not know if the theory governing the strong gravitational field is GR or ECG (assuming that one or the other is the empirically correct theory). However the correct theory must (for a given set of parameters) have the same value of $\beta$ at both intersection points, allowing for its determination.

In certain situations the distance to the source (and hence the value of $\mathcal{D}$) may not
be known. Let us clarify the problem with an example: GR with $\mathcal{D}=0.49972$ yields the
same lines for the image positions as ECG with $\mathcal{D}=0.5$ (the solid blue curves in
Fig.~\ref{fig:finding_source}). In other words,  although $\beta$ can be distinguished via the intersection points of the  $\theta_{p}$ and $|\theta_{s}|$ curves with observation, this is  insufficient to determine $\mathcal{D}$ and distinguish between GR and ECG. In this case a measurement of the differential time delay could be used to break this degeneracy.  In the bottom plot in  Fig.~\ref{fig:finding_source}, we see that the differential time delay between the secondary and primary images in GR with $\mathcal{D}=0.49972$ is bigger than that in ECG with $\mathcal{D}=0.5$. Provided  the source is pulsating (or has otherwise reliable variability), we could measure the differential time delay $t_{d,obs}$. Now, either 
$\bar{t}_{d,{\rm GR}} - t_{d,obs}$
or $t_{d,{\rm ECG}} - t_{d,obs}$
should be zero at a value of $\beta$ consistent with the aforementioned image observations (if not, then both theories would be empirically discredited).
In conjunction with an observation of the primary and secondary images, a time delay measurement can provide enough information to obtain $\beta$ and $\mathcal{D}$ and distinguish the governing theory of the gravitational field of the black hole.
	 
GR and ECG results for magnifications, and the time delays of first and second order relativistic images are, respectively, presented in Tables \ref{tab:Tableiii} and \ref{tab:Tableiv}. First (Second) order relativistic images are produced after the light winds, once (twice) around the black hole before reaching the observer \cite{Ellis}. The angular position of relativistic images $\theta_{1p}$, $|\theta_{1s}|$, $\theta_{2p}$, and $|\theta_{2s}|$ are almost independent of angular source positions. In ECG their values are about $0.2$ nanoarcseconds more than their corresponding values in GR, an effect  too tiny to be observed with today's telescopes, especially with the problem that these relativistic images are highly demagnified. However once they could be observed,   (differential) time delay of relativistic images could be used to test ECG  because their increase compared to GR, as can be seen from Tables \ref{tab:Tableiii} and \ref{tab:Tableiv}.

In Table \ref{tab:Table5} we have studied primary and secondary images in ECG when the source is closer to Sgr A*. In particular, we have taken ${\cal D}=0.05$. Comparing this table with Table \ref{tab:Tablei} (in which ${\cal D}=0.5$), shows that when the source-lens distance is smaller, primary and secondary images get closer to the line of sight to the lens ($\theta_{p}$ and $|\theta_{s}|$ get smaller).  Furthermore, a comparison of Tables \ref{tab:Table5} and \ref{tab:Tableii} shows that the magnification $\mu_{p}$ and $|\mu_{s}|$ and the time delay of the primary image are smaller in the case of ${\cal D}=0.05$ compared to ${\cal D}=0.5$. However the differential time delay $t_d=\tau_s-\tau_p$ is larger in the former case. Similar results hold when the governing theory of gravity is GR~\cite{Virbhadra}.

\begin{figure}
	\centering
	\includegraphics[width=0.5\textwidth]{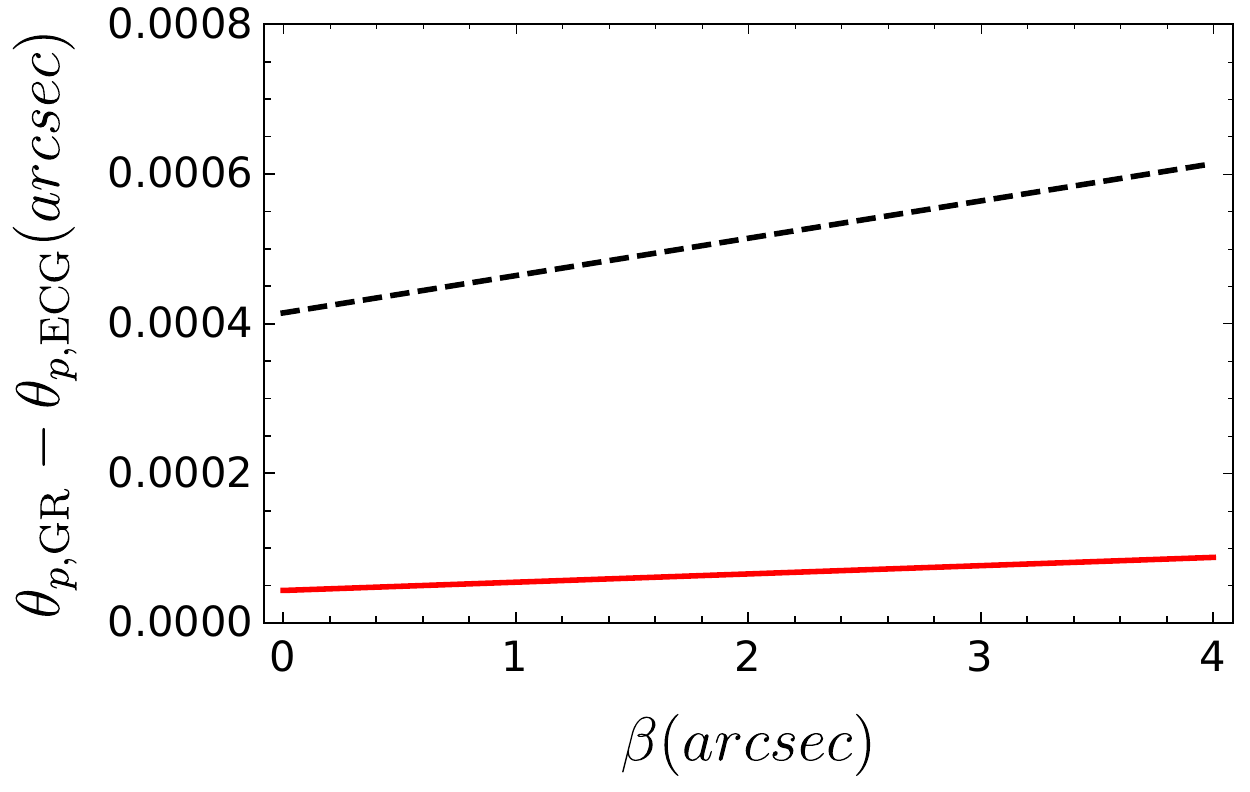}
	\caption{{\bf Deviation of primary image angular position in ECG from GR for Sgr A*}: The deviation increase with angular source position $\beta$. The black dashed line is for the case ${\cal D}=0.5$ and the red line is for ${\cal D}=0.05$. It is obvious that, for a fixed lens-observer distance, the deviation of ECG results for angular positions of primary images from that of GR is larger for sources further away from the lens.}
	\label{fig:diff_theta}
\end{figure}

\bigskip
\begin{figure}
	\centering
	\includegraphics[width=0.5\textwidth]{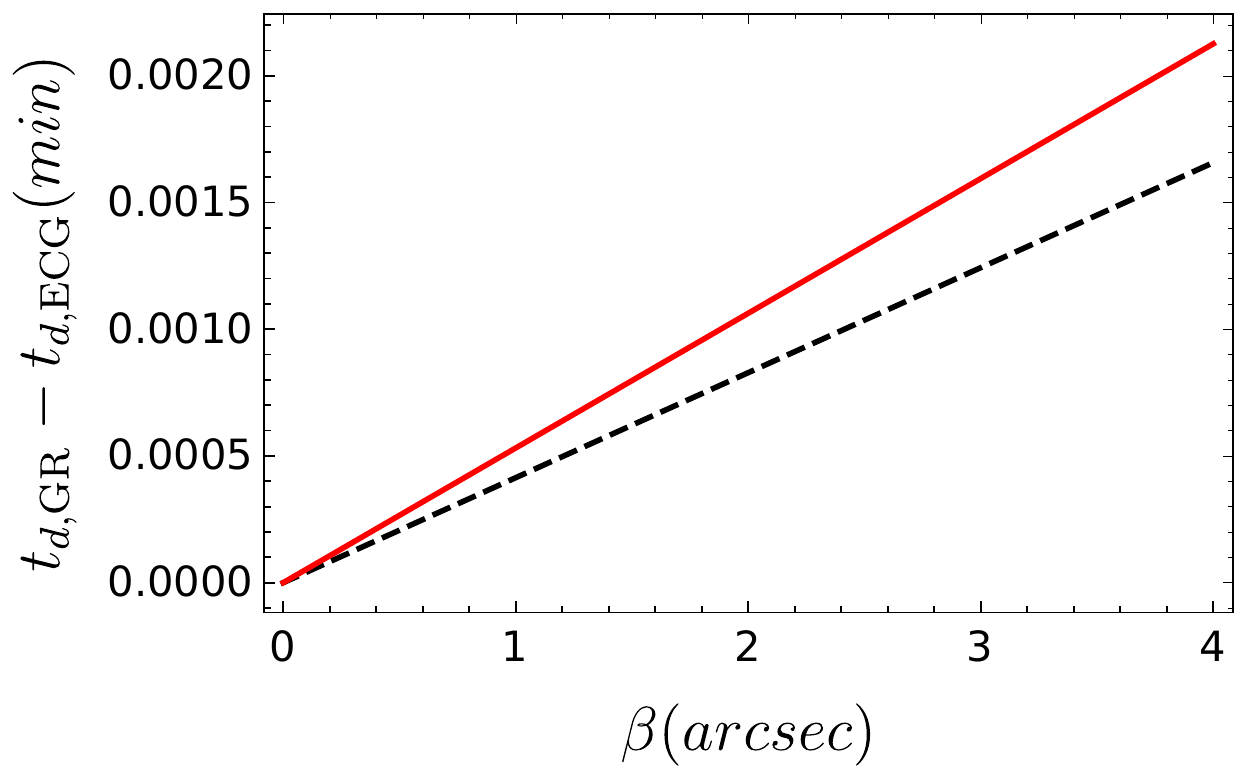}
	\caption{{\bf Deviation of differential time delay in ECG from GR for Sgr A*}: The differential time delay $t_d=\tau_s-\tau_p$ deviate from its corresponding value in GR if ECG governs the strong gravitational field around the black hole. The deviation increase with angular source position $\beta$. The black dashed line is for the case ${\cal D}=0.5$ and the red line is for ${\cal D}=0.05$. Here we   see that, for a fixed lens-observer distance, the deviation of ECG results for the differential time delay from that of GR is larger for sources if the source is closer to the lens.}
	\label{fig:diff_t}
\end{figure}

Although we have not in Table \ref{tab:Table5} presented the corresponding results in GR, we have given relevant comparisons between GR and ECG in Figs.~\ref{fig:diff_theta} and \ref{fig:diff_t}. In Fig.~\ref{fig:diff_theta} we can see that the difference between the results of ECG and GR for the angular positions of primary images is larger in the case ${\cal D}=0.5$. On the other hand, as shown in Fig.~\ref{fig:diff_t}, the deviation of the differential time delay $t_d$ in ECG from its corresponding GR value is larger for ${\cal D}=0.05$.

\begin{figure}
	\centering
	\includegraphics[width=0.5\textwidth]{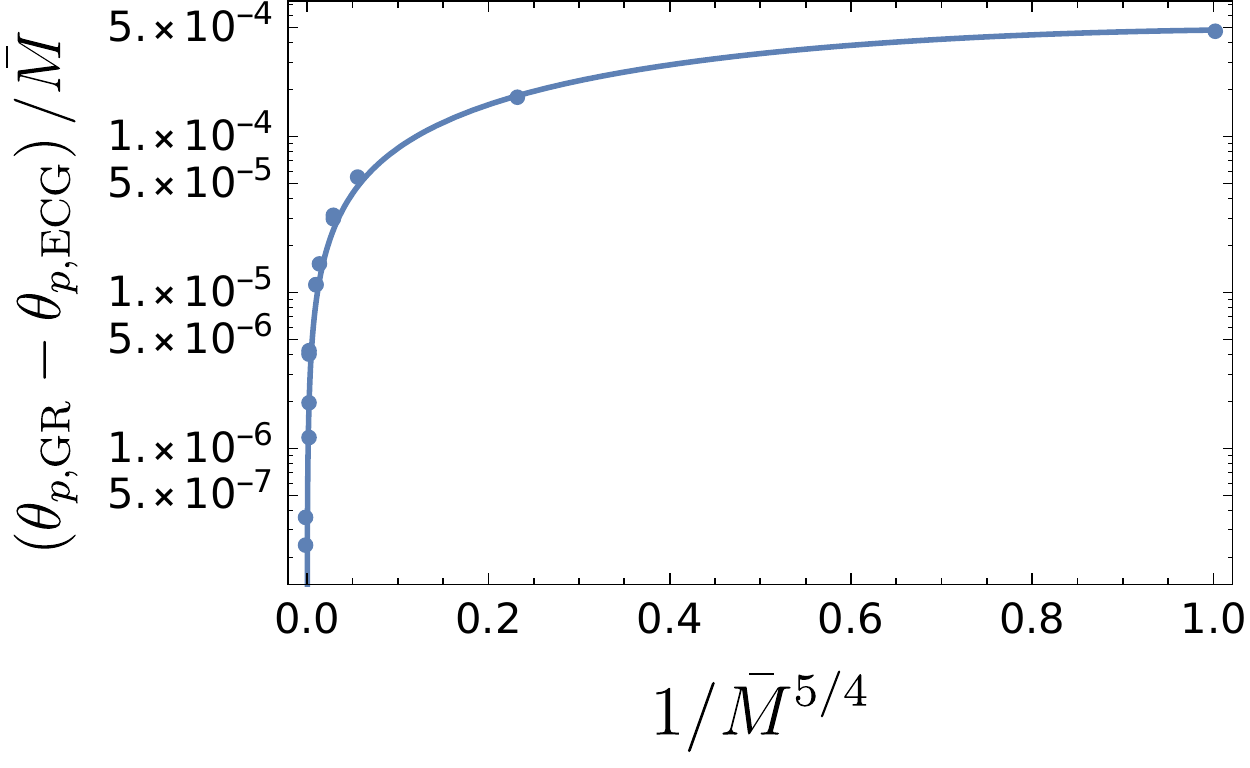}
	\caption{{\bf Deviation of primary image angular position in ECG from GR for different SMBHs}: The ratio $\left(\theta_{p,{\rm GR}}-\theta_{p,{\rm ECG}}\right)/\bar{M}$ increases as $\bar{M}$ decreases. Here $\bar{M}=M/M_{Sgr A*}$, where $M$ is the mass of the SMBH from Table~\ref{tab:Table6}. We have taken ${\cal D}=0.5$.  The dots refer to the numerical results of Table \ref{tab:Table7} for the 14 SMBHs, and  the solid curve is the interpolation between the points.}
	\label{fig:smbh_theta}
\end{figure}

\begin{figure}
	\centering
	\includegraphics[width=0.5\textwidth]{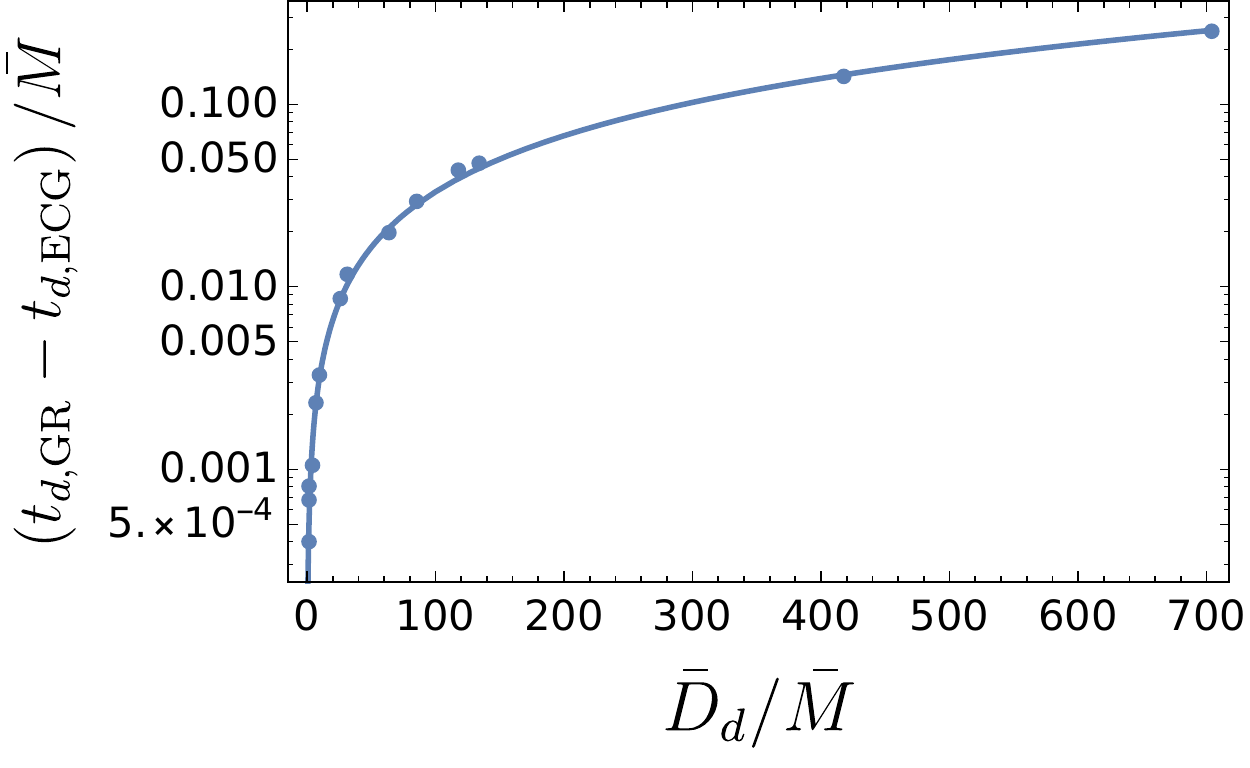}
	\caption{{\bf Deviation of differential time delay in ECG from GR for different SMBHs}: The ratio $\left(t_{d,{\rm GR}}-t_{d,{\rm ECG}}\right)/\bar{M}$ increases as $\bar{D}_d/\bar{M}$ increases. Here $\bar{D}_d= D_d/D_{d,Sgr A*}$ and $\bar{M}=M/M_{Sgr A*}$, where $D_d$ and $M$ are the mass and distance of the SMBH from Table~\ref{tab:Table6}. We have taken ${\cal D}=0.5$.  The dots refer to the numerical results of Table \ref{tab:Table7} for the 14 SMBHs, and   the solid curve is the interpolation between the points.}
	\label{fig:smbh_t}
\end{figure}

We close this section by considering SMBHs in other galaxies.  We wish to see how the GR and ECG predictions for GL differ when the mass and distance of the black hole change from that of  Sgr A*. In Table \ref{tab:Table6} we have collected some updated data of 14 galaxies \cite{mnd,kormendy2013}. We have used these data in Table \ref{tab:Table7} to calculate angular positions and the time delays of primary images in GR as well as ECG, along with the differential time delay $t_d$ between the secondary and primary images for each. 
We have shown   in Fig.~\ref{fig:smbh_theta} how the difference in the angular position of
the primary image between  GR and ECG depends on the mass of the black hole. Differential time delays likewise have a complicated dependence on the black hole mass;  we illustrate this in Fig.~\ref{fig:smbh_t}, where we note that this quantity is less sensitive to the mass and mostly depends on the distance $\bar{D}_d$.

\bigskip

\bigskip

\begingroup
\begin{table*}
	\caption{
		{\bf Image positions and deflection angles of primary and secondary images due to lensing by Sgr A* 
		with ${\cal D}=0.5$}: GR and ECG predictions for angular positions $\theta$ and bending angles $\hat{\alpha}$ are given for different values of angular source position $\beta$. {\bf (a)} $p$ and $s$ refer to primary and secondary images, respectively. {\bf (b)} All angles are in {\em arcseconds}. {\bf (c)} We have used $M_{Sgr A*}=5.94\times 10^9 \, {\rm m}$, $D_d=2.43\times 10^{20} \, {\rm m}$, and $\lambda/M_{Sgr A*}^{4}\approx 1.76\times 10^{-4}$.
	}\label{tab:Tablei}
	%%%%%%%%
	\begin{ruledtabular}
		\begin{tabular}{l cccc cccc}
			%%%%%%%%
			\multicolumn{1}{c}{$\beta$}&
			\multicolumn{4}{c}{General relativity}&
			\multicolumn{4}{c}{Einsteinian Cubic Gravity}\\
			%%%%%%%%%%%%%%%%%%%%%%%%%%%%%%%%%%%%%
&$\theta_{p,{\rm GR}} $&$\hat{\alpha}_{p,{\rm GR}}$&$\theta_{s,{\rm GR}}$&$\hat{\alpha}_{s,{\rm GR}}$&$\theta_{p,{\rm ECG}}$&$\hat{\alpha}_{p,{\rm ECG}}$&$\theta_{s,{\rm ECG}}$&$\hat{\alpha}_{s,{\rm ECG}}$\\
			\hline
			%%%%%%%%%%%%%%%%%%%%%%%%%%%%%%%%%%%%% 
$0	 	 $&$  1.44324     $&$	2.88648    		$&$  -1.44324  $&$    	  2.88648     $&$  1.44284     $&$	  2.88568   	  $&$ -1.44284     $&$    2.88568     	   $\\
			$10^{-3} $&$  1.44374	  $&$   2.88548    		$&$  -1.44274  $&$	   	  2.88748     $&$  1.44334	   $&$    2.88468   	  $&$ -1.44234	   $&$	   2.88668        $\\
			$10^{-2} $&$  1.44825	  $&$   2.87650    		$&$  -1.43825  $&$		  2.89650     $&$  1.44785	   $&$    2.87569   	  $&$ -1.43785     $&$	   2.89570        $\\
			$10^{-1} $&$  1.49411	  $&$   2.78821    		$&$  -1.39411  $&$		  2.98821     $&$  1.49369	   $&$    2.78739   	  $&$ -1.39372     $&$	   2.98743      $\\
			$1	 	 $&$  2.02740	  $&$   2.05479    		$&$  -1.02740  $&$    	  4.05480     $&$  2.02692	   $&$    2.05384   	  $&$ -1.02711     $&$	   4.05422          $\\
			$2	 	 $&$  2.75583	  $&$   1.51166    		$&$  -0.755838 $&$    	  5.51167     $&$  2.75534	   $&$    1.51069   	  $&$ -0.755650    $&$    5.51130         $\\
			$3	 	 $&$  3.58157	  $&$   1.16314    		$&$  -0.581575 $&$		  7.16322     $&$  3.58095	   $&$    1.16190   	  $&$ -0.581472    $&$	   7.16294         $\\
			$4	 	 $&$  4.46636	  $&$   0.932720   		$&$  -0.466372 $&$    	  8.93274     $&$  4.46578	   $&$    0.931568  	  $&$ -0.466319	   $&$    8.93264        $\\
			%%%%%%%%%%%%%%%%%
		\end{tabular}
	\end{ruledtabular}
\end{table*}
\endgroup

\begingroup
\begin{table*}
	\caption{
		{\bf Magnifications and time delays of primary and secondary images due to lensing by Sgr A* with ${\cal D}=0.5$}: GR and ECG predictions for magnifications $\mu$, time delays $\tau$, and differential time delays $t_d=\tau_s -\tau_p$ are given for different values of angular source position $\beta$. {\bf (a)} As in Table \ref{tab:Tablei}. {\bf (b)} $\beta$ is in {\em arcseconds} and time delays are in {\em minutes}. {\bf (c)} As in Table \ref{tab:Tablei}.
	}\label{tab:Tableii}
	%%%%%%%%
	\begin{ruledtabular}
		\begin{tabular}{l cccc cccc}
			%%%%%%%%
			\multicolumn{1}{c}{$\beta$}&
			\multicolumn{4}{c}{General relativity}&
			\multicolumn{4}{c}{Einsteinian Cubic Gravity}\\
			%%%%%%%%%%%%%%%%%%%%%%%%%%%%%%%%%%%%%
					  &$\mu_{p,{\rm GR}} $&$\tau_{p,{\rm GR}}$&$\mu_{s,{\rm GR}}$&$t_{d,{\rm GR}}$&$\mu_{p,{\rm ECG}}$&$\tau_{p,{\rm ECG}}$&$\mu_{s,{\rm ECG}}$&$t_{d,{\rm ECG}}$\\
			\hline
			%%%%%%%%%%%%%%%%%%%%%%%%%%%%%%%%%%%%% 
			$0	 	 $&$	\times    $&$	16.588179 $&$ \times      $&$   0            $&$	\times    $&$	16.588916 $&$ \times      $&$   0   	        $\\
			$10^{-3} $&$	722.117   $&$	16.587254 $&$ -721.117    $&$	0.001830     $&$	721.715   $&$	16.588025 $&$ -720.715    $&$	0.001829        $\\
			$10^{-2} $&$	72.6630   $&$	16.579045 $&$ -71.6630    $&$	0.018297     $&$	72.6226   $&$	16.579774 $&$ -71.6230    $&$	0.018295        $\\
			$10^{-1} $&$	7.72915   $&$	16.498249 $&$ -6.72916    $&$	0.183015     $&$	7.72496   $&$	16.499004 $&$ -6.72531    $&$	0.182974      $\\
			$1	 	 $&$	1.34553   $&$	15.813781 $&$ -0.345536   $&$	1.865927     $&$	1.34500   $&$	15.814701 $&$ -0.345307   $&$	1.865259          $\\
			$2	 	 $&$	1.08134   $&$	15.254996 $&$ -0.0813405  $&$	3.934199     $&$	1.08107   $&$	15.256052 $&$ -0.0812885  $&$	3.933519         $\\
			$3	 	 $&$	1.02708   $&$	14.835070 $&$ -0.0270804  $&$	6.358834     $&$	1.02685   $&$	14.836730 $&$ -0.0270673  $&$	6.357461         $\\
			$4	 	 $&$ 	1.01102   $&$ 	14.505271 $&$ -0.0110231  $&$ 	9.236943     $&$ 	1.01087   $&$ 	14.507129 $&$ -0.0110196  $&$ 	9.235372        $\\
			%%%%%%%%%%%%%%%%%
		\end{tabular}
	\end{ruledtabular}
\end{table*}
\endgroup

\begingroup
\begin{table*}
	\caption{
		{\bf Magnifications and time delays of first order relativistic images due to lensing by Sgr A* with ${\cal D}=0.5$}: GR and ECG predictions for magnifications $\mu$ and time delays $\tau$ are given for different values of angular source position $\beta$. {\bf (a)} $1p$ and $1s$ refer to first order relativistic images on the same side as primary and secondary images, respectively. {\bf (b)} As in Table \ref{tab:Tableii}. {\bf (c)} As in Table \ref{tab:Tablei}. {\bf (d)} Angular positions of first order relativistic images in GR and ECG are, respectively, $\theta_{1p,{\rm GR}}\approx -\theta_{1s,{\rm GR}}\approx 26.2691 \mu as$ and $\theta_{1p,{\rm ECG}}\approx -\theta_{1s,{\rm ECG}}\approx 26.2693 \mu as$ and are highly insensitive to the angular source position $\beta$.
	}\label{tab:Tableiii}
	%%%%%%%%
	\begin{ruledtabular}
		\begin{tabular}{l cccc cccc}
			%%%%%%%%
			\multicolumn{1}{c}{$\beta$}&
			\multicolumn{4}{c}{General relativity}&
			\multicolumn{4}{c}{Einsteinian Cubic Gravity}\\
			%%%%%%%%%%%%%%%%%%%%%%%%%%%%%%%%%%%%%
					   &$\mu_{1p,{\rm GR}} $&$\tau_{1p,{\rm GR}}$&$\mu_{1s,{\rm GR}}$&$\tau_{1s,{\rm GR}}$&$\mu_{1p,{\rm ECG}}$&$\tau_{1p,{\rm ECG}}$&$\mu_{1s,{\rm ECG}}$&$\tau_{1s,{\rm ECG}}$\\
			\hline
			%%%%%%%%%%%%%%%%%%%%%%%%%%%%%%%%%%%%% 
			$0        $&$	 \times	 		 $&$ 42.673253 $&$	   \times 		   $&$ 42.673253 $&$   \times		    $&$ 42.673306 $&$	  \times		  $&$ 42.673306         $\\
			$10^{-6}  $&$8.43\times 10^{-12} $&$ 42.673253 $&$-8.43\times 10^{-12} $&$ 42.673253 $&$8.42\times 10^{-12} $&$ 42.673306 $&$-8.42\times 10^{-12} $&$ 42.673306  		$\\
			$10^{-5}  $&$8.43\times 10^{-13} $&$ 42.673253 $&$-8.43\times 10^{-13} $&$ 42.673253 $&$8.42\times 10^{-13} $&$ 42.673306 $&$-8.42\times 10^{-13} $&$ 42.673306  		$\\
			$10^{-4}  $&$8.43\times 10^{-14} $&$ 42.673253 $&$-8.43\times 10^{-14} $&$ 42.673253 $&$8.42\times 10^{-14} $&$ 42.673306 $&$-8.42\times 10^{-14} $&$ 42.673306  		$\\
			$10^{-3}  $&$8.43\times 10^{-15} $&$ 42.673256 $&$-8.43\times 10^{-15} $&$ 42.673256 $&$8.42\times 10^{-15} $&$ 42.673308 $&$-8.42\times 10^{-15} $&$ 42.673308 		$\\
			$10^{-2}  $&$8.43\times 10^{-16} $&$ 42.673280 $&$-8.43\times 10^{-16} $&$ 42.673280 $&$8.42\times 10^{-16} $&$ 42.673337 $&$-8.42\times 10^{-16} $&$ 42.673337  		$\\
			$10^{-1}  $&$8.43\times 10^{-17} $&$ 42.676417 $&$-8.43\times 10^{-17} $&$ 42.676420 $&$8.42\times 10^{-17} $&$ 42.676474 $&$-8.42\times 10^{-17} $&$ 42.676477  		$\\
			$1        $&$8.43\times 10^{-18} $&$ 42.990190 $&$-8.43\times 10^{-18} $&$ 42.990224 $&$8.42\times 10^{-18} $&$ 42.990244 $&$-8.42\times 10^{-18} $&$ 42.990275  		$\\
			%%%%%%%%%%%%%%%%%
		\end{tabular}
	\end{ruledtabular}
\end{table*}
\endgroup

\begingroup
\begin{table*}
	\caption{
		{\bf Magnifications and time delays of second order relativistic images due to lensing by Sgr A* with ${\cal D}=0.5$}: GR and ECG predictions for magnifications $\mu$, time delays $\tau$, and differential time delays $\tau_{2p}-\tau_{1p}$ are given for different values of angular source position $\beta$. {\bf (a)} $2p$ and $2s$ refer to second order relativistic images on the same side as primary and secondary images, respectively. {\bf (b)} As in Table \ref{tab:Tableii}. {\bf (c)} As in Table \ref{tab:Tablei}. {\bf (d)} Angular positions of second order relativistic images in GR and ECG are, respectively, $\theta_{2p,{\rm GR}}\approx -\theta_{2s,{\rm GR}}\approx 26.2362 \mu as$ and $\theta_{2p,{\rm ECG}}\approx -\theta_{2s,{\rm ECG}}\approx 26.2364 \mu as$ and are highly insensitive to the angular source position $\beta$. {\bf (e)} $\mu_{2s}= -\mu_{2p}$ to a very good approximation. {\bf (f)} Explicit values of $\tau_{1p}$ are given in Table \ref{tab:Tableiii}.
	}\label{tab:Tableiv}
	%%%%%%%%
	\begin{ruledtabular}
		\begin{tabular}{l cccc cccc}
			%%%%%%%%
			\multicolumn{1}{c}{$\beta$}&
			\multicolumn{4}{c}{General relativity}&
			\multicolumn{4}{c}{Einsteinian Cubic Gravity}\\
			%%%%%%%%%%%%%%%%%%%%%%%%%%%%%%%%%%%%%
			&$\mu_{2p,{\rm GR}}$&$\tau_{2p,{\rm GR}}$&$\tau_{2s,{\rm GR}}$&$(\tau_{2p}-\tau_{1p})_{{\rm GR}}$&$\mu_{2p,{\rm ECG}}$&$\tau_{2p,{\rm ECG}}$&$\tau_{2s,{\rm ECG}}$&$(\tau_{2p}-\tau_{1p})_{{\rm ECG}}$\\
			\hline
			%%%%%%%%%%%%%%%%%%%%%%%%%%%%%%%%%%%%% 
			$0        $&$     \times              $&$ 53.452474 $&$ 53.452474 $&$ 10.779221$&$     \times              $&$ 53.452590 $&$ 53.452590 $&$ 10.779284       $\\
			$10^{-6}  $&$     1.44\times 10^{-14} $&$ 53.452474 $&$ 53.452474 $&$ 10.779221$&$     9.45\times 10^{-15} $&$ 53.452590 $&$ 53.452590 $&$ 10.779284  		$\\
			$10^{-5}  $&$     1.44\times 10^{-15} $&$ 53.452474 $&$ 53.452474 $&$ 10.779221$&$     9.45\times 10^{-16} $&$ 53.452590 $&$ 53.452590 $&$ 10.779284  		$\\
			$10^{-4}  $&$     1.44\times 10^{-16} $&$ 53.452474 $&$ 53.452474 $&$ 10.779221$&$     9.45\times 10^{-17} $&$ 53.452590 $&$ 53.452590 $&$ 10.779284  		$\\
			$10^{-3}  $&$     1.44\times 10^{-17} $&$ 53.452477 $&$ 53.452477 $&$ 10.779221$&$     9.45\times 10^{-18} $&$ 53.452592 $&$ 53.452592 $&$ 10.779284 		$\\
			$10^{-2}  $&$     1.44\times 10^{-18} $&$ 53.452502 $&$ 53.452502 $&$ 10.779221$&$     9.45\times 10^{-19} $&$ 53.452621 $&$ 53.452621 $&$ 10.779284  		$\\
			$10^{-1}  $&$     1.44\times 10^{-19} $&$ 53.455638 $&$ 53.455642 $&$ 10.779221$&$     9.45\times 10^{-20} $&$ 53.455758 $&$ 53.455759 $&$ 10.779284  		$\\
			$1        $&$     1.44\times 10^{-20} $&$ 53.769411 $&$ 53.769445 $&$ 10.779221$&$     9.45\times 10^{-21} $&$ 53.769528 $&$ 53.769557 $&$ 10.779284  		$\\
			%%%%%%%%%%%%%%%%%
		\end{tabular}
	\end{ruledtabular}
\end{table*}
\endgroup

\begingroup
\begin{table*}
	\caption{
		{\bf Primary and secondary images due to lensing by Sgr A* in ECG with ${\cal D}=0.05$}: Angular  positions $\theta$, bending angles $\hat{\alpha}$, magnifications $\mu$, time delays $\tau$, and the differential time delay $t_d=\tau_s-\tau_p$ are given for different values of angular source position $\beta$. {\bf (a)} As in Table \ref{tab:Tablei}. {\bf (b)} All angles are in {\em arcseconds} and time delays are in {\em minutes}. {\bf (c)} As in Table \ref{tab:Tablei}.
	}\label{tab:Table5}
	%%%%%%%%
	\begin{ruledtabular}
		\begin{tabular}{l cccc cccc}  
			%%%%%%%%%%%%%%%%%%%%%%%%%%%%%%%%%%%%%
			$\beta	 $&$  \theta_p    $&$\hat{\alpha}_p$&$   \mu_p    $&$  \tau_p     $&$ \theta_{s}$&$\hat{\alpha}_{s}$&$  \mu_{s}    $&$  t_d			   $\\
			\hline
			%%%%%%%%%%%%%%%%%%%%%%%%%%%%%%%%%%%%%
			$0	 	 $&$  0.45636     $&$	9.12727    $&$	\times    $&$	16.164666 $&$ -0.45636  $&$    9.12727     $&$ \times      $&$  0    		   $\\
			$10^{-3} $&$  0.45686	  $&$   9.11727    $&$	228.661   $&$	16.161775 $&$ -0.45586  $&$	   9.13727     $&$ -227.661    $&$	0.005786       $\\
			$10^{-2} $&$  0.46139	  $&$   9.02780    $&$	23.3201   $&$	16.135380 $&$ -0.45139  $&$	   9.22782     $&$ -22.3202    $&$	0.058370       $\\
			$10^{-1} $&$  0.50909	  $&$   8.18178    $&$	2.82242   $&$	15.890648 $&$ -0.40910  $&$	   10.1820     $&$ -1.82255    $&$	0.579733       $\\
			$1	 	 $&$  1.17691	  $&$   3.53816    $&$	1.02306   $&$	14.352152 $&$ -0.17698  $&$	   23.5394     $&$ -0.02313    $&$	6.792566       $\\
			$2	 	 $&$  2.09915	  $&$   1.98296    $&$	1.00220   $&$	13.521637 $&$ -0.09923  $&$    41.9847     $&$ -0.00224    $&$	17.96494       $\\
			$3	 	 $&$  3.06782	  $&$   1.35631    $&$	1.00046   $&$	13.004608 $&$ -0.06791  $&$	   61.3581     $&$ -0.00049    $&$	34.84736       $\\
			$4	 	 $&$  4.05133	  $&$   1.02664    $&$ 	1.00014   $&$ 	12.631702 $&$ -0.05143  $&$    81.0285     $&$ -0.00016    $&$ 	57.78042       $\\
			%%%%%%%%%%%%%%%%%
		\end{tabular}
	\end{ruledtabular}
\end{table*}
\endgroup

\begingroup
\begin{table*} 
	\caption{ 
		{\bf Masses and distances of SMBHs}: Masses ($M$) and distances ($D_d$) of SMBHs at the center of 14 galaxies. The data for Sgr A* at the center of Milky Way Galaxy has been taken from~\cite{mnd}. The data of other black holes are from~\cite{kormendy2013}.
	}\label{tab:Table6}
	%%%%%%%%
	\begin{ruledtabular}
		\begin{tabular}{cccc cccc}  
			%%%%%%%%%%%%%%%%%%%%%%%%%%%%%%%%%%%%% 
			Galaxy   &           $M$ (m)      &          $D_d$ (m)   &     $D_d/M$        & Galaxy   &         $M$ (m)      &       $D_d$ (m)      &      $D_d/M$         \\
			\hline
			%%%%%%%%%%%%%%%%%%%%%%%%%%%%%%%%%%%%%
			Milky Way&$  5.94\times 10^9	 $&$ 2.43\times 10^{20} $&$4.09\times 10^{10}$& M31      &$ 2.11\times 10^{11} $&$ 2.39\times 10^{22} $&$1.13\times 10^{11}$     \\
			M87      &$  9.08\times 10^{12}  $&$ 5.15\times 10^{23} $&$5.67\times 10^{10}$& NGC 1023 &$ 6.10\times 10^{10} $&$ 3.34\times 10^{23} $&$5.48\times 10^{12}$      \\
			NGC 1194 &$  1.05\times 10^{11}  $&$ 1.79\times 10^{24} $&$1.70\times 10^{13}$& NGC 1316 &$ 2.50\times 10^{11} $&$ 6.47\times 10^{23} $&$2.59\times 10^{12}$      \\
			NGC 1332 &$  2.17\times 10^{12}  $&$ 6.99\times 10^{23} $&$3.22\times 10^{11}$& NGC 1407 &$ 6.87\times 10^{12} $&$ 8.95\times 10^{23} $&$1.30\times 10^{11}$     \\
			NGC 3607 &$  2.02\times 10^{11}  $&$ 6.99\times 10^{23} $&$3.46\times 10^{12}$& NGC 3608 &$ 6.87\times 10^{11} $&$ 7.02\times 10^{23} $&$1.02\times 10^{12}$    \\
			NGC 4261 &$  7.81\times 10^{11}  $&$ 9.99\times 10^{23} $&$1.28\times 10^{12}$& NGC 4374 &$ 1.37\times 10^{12} $&$ 5.71\times 10^{23} $&$4.17\times 10^{11}$      \\
			NGC 4382 &$  1.92\times 10^{10}  $&$ 5.52\times 10^{23} $&$2.88\times 10^{13}$& NGC 4459 &$ 1.03\times 10^{11} $&$ 4.94\times 10^{23} $&$4.80\times 10^{12}$      \\
			%%%%%%%%%%%%%%%%%%%%%%%%%%%%%%%%%%%%%
		\end{tabular}
	\end{ruledtabular}
\end{table*}
\endgroup

\begingroup
\begin{table*}
	\caption{
		{\bf Image positions and time delays due to lensing by SMBHs}: GR and ECG predictions for angular positions $\theta$ and the time delays $\tau$ of primary images as well as the differential time delays $t_d=\tau_s-\tau_p$ are given for different SMBHs. We have also presented the difference between GR and ECG predictions $\theta_p$ and $t_d$. {\bf (a)} As in Table \ref{tab:Tablei}. {\bf (b)} All angles are in {\em arcseconds} and time delays are in {\em minutes}. {\bf (c)} We have taken ${\cal D}=0.5$, $\beta =1 arcsecond$, and $\lambda \approx 2.19\times 10^{35}{\rm m}^4$.
	}\label{tab:Table7}
	%%%%%%%%
	\begin{ruledtabular}
		\begin{tabular}{l ccc ccccc}
			%%%%%%%%
			\multicolumn{1}{c}{Galaxy}&
			\multicolumn{3}{c}{General relativity}&
			\multicolumn{5}{c}{Einsteinian Cubic Gravity}\\
			%%%%%%%%%%%%%%%%%%%%%%%%%%%%%%%%%%%%%
			&$\theta_{p,{\rm GR}}$&$  \tau_{p,{\rm GR}}  $&$  t_{d,{\rm GR}}  $&$ \theta_{p,{\rm ECG}} $&$ \tau_{p,{\rm ECG}} $&$  t_{d,{\rm ECG}}  $&$\theta_{p,{\rm GR}}-\theta_{p,{\rm ECG}}$&$t_{d,{\rm GR}}-t_{d,{\rm ECG}}$\\
			\hline
			%%%%%%%%%%%%%%%%%%%%%%%%%%%%%%%%%%%%%
			Milky Way &$  2.02740    $&$   15.813781   $&$  1.865927  $&$  	 2.02692     $&$  15.814701   $&$  1.865259   $&$  	   	0.00048			     $&$    0.000668      $\\
			M31       &$  1.50121    $&$   572.58323   $&$  114.0235  $&$  	 1.50082     $&$  572.62936   $&$  113.9949   $&$  	   	0.00039			     $&$    0.028667      $\\
			M87       &$  1.82348    $&$   24351.162   $&$  3386.221  $&$  	 1.82311     $&$  24352.391   $&$  3385.616   $&$  	   	0.00037			     $&$    0.605146      $\\
			NGC 1023  &$  1.01532    $&$   168.51053   $&$  506.3728  $&$  	 1.01478     $&$  168.99269   $&$  505.8833   $&$  	   	0.00055			     $&$    0.489513      $\\
			NGC 1194  &$  1.00495    $&$   288.99871   $&$  2485.346  $&$  	 1.00442     $&$  291.45661   $&$  2482.897   $&$  	   	0.00053			     $&$    2.449040      $\\
			NGC 1316  &$  1.03184    $&$   689.09579   $&$  1091.665  $&$  	 1.03136     $&$  689.95441   $&$  1090.821   $&$  	   	0.00048			     $&$    0.844443      $\\
			NGC 1332  &$  1.21708    $&$   5949.7675   $&$  2142.847  $&$  	 1.21665     $&$  5950.8862   $&$  2141.989   $&$  	   	0.00043			     $&$    0.858462      $\\
			NGC 1407  &$  1.45027    $&$   18653.830   $&$  4008.966  $&$  	 1.44985     $&$  18655.784   $&$  4007.738   $&$  	   	0.00042		         $&$    1.228134      $\\
			NGC 3607  &$  1.02405    $&$   558.79222   $&$  1126.096  $&$  	 1.02354     $&$  559.78355   $&$  1125.090   $&$  	   	0.00052		         $&$    1.005752      $\\
			NGC 3608  &$  1.07727    $&$   1892.5635   $&$  1461.592  $&$  	 1.07678     $&$  1893.6209   $&$  1460.614   $&$  	   	0.00049		         $&$    0.978089      $\\
			NGC 4261  &$  1.06265    $&$   2154.3003   $&$  1960.302  $&$  	 1.06211     $&$  2155.8850   $&$  1958.763   $&$  	   	0.00054		         $&$    1.539308      $\\
			NGC 4374  &$  1.17344    $&$   3750.3521   $&$  1586.191  $&$  	 1.17299     $&$  3751.3033   $&$  1585.447   $&$  	   	0.00045		         $&$    0.744141      $\\
			NGC 4382  &$  1.00295    $&$   53.070206   $&$  750.1476  $&$  	 1.00239     $&$  53.891505   $&$  749.3269   $&$  	   	0.00057		         $&$    0.820750      $\\
			NGC 4459  &$  1.01740    $&$   283.95236   $&$  761.0761  $&$  	 1.01685     $&$  284.69833   $&$  760.3394   $&$  	   	0.00055		         $&$    0.736695      $\\
			%%%%%%%%%%%%%%%%%
		\end{tabular}
	\end{ruledtabular}
\end{table*}
\endgroup

\section{Concluding remarks}
\label{sec:con}

In its predictions for GL due to SMBHs,   ECG exhibits small but potentially observable departures from GR.
Taking the ECG coupling constant  to be $\lambda = 4.57 \times 10^{22} M_{\astrosun}^4$, for which  ECG  passes all   Solar System tests to date~\cite{OURS}, we find that the angular positions of primary and secondary images deviate from that of GR by an amount of order of miliarcseconds.  The ECG results for the differential time delay, associated with primary and secondary images, could be some tenths of seconds shorter than the results of GR.

It is important to note that for the primary/secondary images to be produced, the light from the source should pass the black hole at a closest distance of order $10^{5}\, r_+$, where $r_+$ is the radius of event horizon. We  have  shown even in this large distance from the black hole that ECG  effects may be observable. One does not have to observe gravitational effects in the vicinity of an horizon to test ECG.

 There are several short period stars (the so-called S-stars) orbiting around Sgr A* whose semimajor axes are less than $10^{5}\, r_+$ \cite{gillessen2017}. Nowadays the observation of these S-stars are possible with good precision \cite{abuter2017}. We propose, as a direction of future study, to investigate the orbit of S-stars in ECG and to compare it with observational results now available \cite{abuter2017,refId0}.

As for GR~\cite{Ellis,Virbhadra}, in ECG relativistic images are produced after the light winds around the black hole. For these images to be produced the light must pass the black hole very closely. Consider the first order relativistic image. The closest approach of the light is $\sim 1.55 \, r_+$, which is very close to the radius of the photon sphere, $r_{ps}=1.5 \, r_+$, where the shadow is produced. The light must get closer and closer to the photon sphere to produce higher and higher order relativistic images.
 We have seen in our previous paper~\cite{OURS} that the effects of ECG on the angular radius of the shadow of Sgr A* is less than 1 nanoarcseconds. Here we see that the same thing is also true for the angular positions of relativistic images. In this case the differential time delay between relativistic images could be used to test ECG, if (since they are highly demagnified) these images could ever be  observed.

 Finally we have also studied GR and ECG predictions for lensing effects by some SMBHs in other galaxies. We find that GR and ECG results for the differential time delay between primary and secondary images could  differ by an amount of more than one minute for some distant SMBHs. The deviation between GR and ECG predictions for image angular positions depends mostly on the mass of black hole and is reminiscent to what we have found in~\cite{OURS}. Very massive ECG black holes are almost like ordinary Schwarzschild black holes. However intermediate mass ECG black holes deviate significantly. This point needs further study and we leave it for future work.

\section*{Acknowledgements}
This work was supported in part by the Natural Sciences and Engineering Research Council of Canada.
 We wish to thank Prof. Yousef Sobouti for comments and helpful discussions concerning this work. MBJP is grateful for the hospitality of IASBS, where part of this work was completed.

\bibliography{mybib}
\end{document}